# Humidity, the Dominating Force of Thermal Updrafts

Oliver Predelli [*]
Braunschweig, Germany
segelflieger@e-mail.de

Ronald Niederhagen [*]
Marzling, Germany
ronald_niederhagen@freenet.de

## Abstract

This paper describes the most significant sources of errors and disturbance when measuring temperature and humidity with a glider. Measurement flights have been started from different airports in Germany, between April and August of 2018, resulting in a collection of over 90 hours of flight data logs. We show how error correction can be applied to the measurement data. Analysis of the data indicates that core assumptions of the theories of thermals, which have been published for decades cannot be backed up by our measurement data. In contrast we present a revised view of temperature and humidity inside thermals. As a result traditional understanding of temperature distribution and entrainment processes must be revised.

## Introduction

It is generally assumed that thermals are warmer than the ambient air. Especially in publications intended for a broad readership, this statement can be found. So on websites, in magazines, but also in literature for glider pilots. Even the term "thermal" implies heat as the driving influence. "The main source for updraft in the summer months are warm bubbles of air that rise when the ground is heated by sunlight" [Dwd19], writes for example the German Weather Service on its website. Statements, such as "for the dynamics it is crucial to know the temperature difference $\Delta T$ between the heated air parcels and the free atmosphere" [Lie93], and "the degree of correlation between temperature and vertical velocity [is significant]" [Mil78] have influenced the collective wisdom of the gliding community over decades.

But is that really true? Different phenomena make us doubt that an air parcel shall only rise because it is warmer than its ambient air. Some well known phenomena cannot be solely described by a temperature advance of thermal updraft air compared to the environment: Why do we predominantly find blue thermals on the lee side of lakes? Why is the cloud-base independent of soil properties? Why can updraft velocity be calculated based on humidity? [Pre17].

One of the best-known formulas for calculating the thermal updraft velocity goes back to James Deardorff, who introduced his convective velocity scale in 1970 [Dea70]:

$$w^* = [\frac{g}{T} z_i \cdot \overline{(w\,T)_0}]^{\frac{1}{3}} \tag{1}$$

He came up with it, because he folded Monin-Obukhov's description of horizontal winds [Mon54] by 90° upwards. And just like Monin-Obukhov he has set a few premises, such as "the deviation of density and temperature from the standard values are proportional" and "the Archimedian force only depends on layer temperature and temperature deviation." In other words: Thermals shall be warmer than the environment. Therefore he used the "T" in his equations. Later scientists replaced the temperature T by the virtual potential temperature $\theta_v$, which seems to be much wiser, as buoyancy is driven by density differences [Stu88] and the updraft speed is thus dependent on both, temperature and humidity:

$$w^* = [\frac{g}{\theta_v} z_i \cdot \overline{w'\theta_v'}]^{\frac{1}{3}} \tag{2}$$

To study the influence of temperature and humidity on thermal updrafts a Ventus-2 glider (**Fig 1**) was equipped with environmental sensors. The data acquisition was linked to the on-board aviation instrument data. In-flight measurements were carried out between April and August 2018 in Germany across plains, low mountain range and foothills of the Alps. Weather conditions varied between clear and overcast, between cool spring and hot summer.

## In-flight measuring equipment

Mobile Devices and Internet of Things (IoT) have triggered a strong investment in sensor technology (Micro-Electro-Mechanical Systems, MEMS) and small portable compute devices (SoC). Over the past decade this investment has led to significant quality improvements as well as cost reduction. It is straight forward to use this equipment in a glider to measure air temperature, humidity and air pressure. Gliders are especially suitable for detailed analysis of thermals because they naturally use thermals flying at low speeds in tight circles.

Nevertheless measuring air temperature and humidity in flight is nontrivial. There are many sources of errors and mistakes which have to be considered in order to understand the validity of the data. Examples of such errors include the heat capacity of the glider's fuselage which may impact the temperature of the air around it. Or a "dry offset" for humidity sensors which were not designed for high airspeed. Also the result of a water-to-air mixing ratio calculation will be wrong if the temperature measurement was erroneous in the first place. A detailed understanding of the causes to these errors is the basis for developing error correction algorithms. They can be used to eliminate errors in a post-process after the data acquisition. Ignoring these erroneous effects may result in faulty interpretation of the data. Several attempts of former in-flight data acquisition suffer from that.

A sensor of type BME280 from Bosch Sensortec [Bos16] is used to sample humidity, pressure and temperature data. The sensor is operated from a dedicated micro controller which records the data from the BME280 together with a GPS-based timestamp (**Fig 2**). The data from the glider's flight and navigation instruments are recorded in a separate unit also together with a GPS-based timestamp. Both units' sample rate is 2 Hz. After the flight the two recorded data streams are synchronized - by means of the GPS-based timestamps - merged together and the result is written to a CSV file containing one data vector for every half second. Among other data the data vector includes humidity, pressure and temperature, as well as Pstat, Ppitot, Pte, GPS-fix, course, TE-vario and airspeed.

It is important to pick the optimal position for the environmental sensors. Unpredictable influences must be avoided or at least minimized and systematic influences must be modeled to allow for data correction in the post-process. Systematic influences include variation in airspeed, influences of parasitic heat capacity upstream or around the sensor, or heat radiation from the sun. Unpredictable influences include turbulent airflow around the sensor or yawing of the glider which changes the airflow around the sensor as the alignment changes between the glider's roll axis and the glider's motion vector. Thus the sensor may be exposed to either fresh air or air which has flown over the surface of the glider.

For the data acquisition of this publication the sensor was positioned inside the ventilation channel to the cockpit of the glider. Thus it is protected from direct sunlight and the misalignment of the glider's roll axis has nearly no effect.

## Temperature correction

### Basic considerations

The result of the effects mentioned above is that the measured temperature suffers from a time lag relative to the temperature of interest. On falling air temperature (during ascend) the sensor reads a slightly warmer temperature. Whereas in rising air temperature (during descend) the reading is slightly colder. Plotting altitude over temperature exhibits a more or less distinct loop form, very much resembling a hysteresis curve (**Fig 3**).

There are some older publications which interpret this loop form as a temperature difference between the thermal and the ambient air with increased potential temperature during ascent and reduced potential temperature during descent. One example where the loop form becomes apparent is the tow phase. During that phase the tow plane pulls the glider to the release point. Typically this is an ascent outside of thermals. There should be no increase of the potential temperature under these conditions. If it was, a time lag was influencing the temperature measurement.

The most noticeable influence on the sensor's operation is caused by the presence of heat capacity upstream from the sensor and the heat capacity of the sensor package and fixture itself (**Fig 4**).

The ambient air with its temperature $T_0$ heats the aircraft and the sensing equipment. There is always a heat transfer $\dot{Q}_1$ from the environment to the aircraft and its equipment. At the same time, there is always a heat transfer $\dot{Q}_2$ from the aircraft and the sensing equipment back into the measurement air: On its way to the temperature sensor, the measurement air sweeps along the surface of the aircraft and through the measuring equipment. The aircraft's surface temperature $T_1$ follows the ambient temperature $T_0$ time-delayed with a first order low-pass filter. This delay is caused by the heat capacity of the aircraft's fuselage. Part of the surface temperature $T_1$ is transferred to the measurement air. It can be shown that the heat transfer from the aircraft back into the measurement air depends on the temperature difference between the aircraft and the environmental air flowing into the measurement equipment. This heat transfer is represented by the factor $f$ in the block diagram, with $0 \leq f \leq 1$. Depending on whether the surface temperature is warmer or colder than the ambient temperature, the measurement air is warmed or cooled. $T_2$ represents the measured air temperature at the sensor position.

It depends on the measuring setup how strong the temperature change of the measurement air is. However, an influence always takes place. At least because of the housing of the probe.

## Physical description of the measuring system

In Fig 4 we show a conceptual picture of the measuring device in proximity of the parts of the fuselage of the glider. In this case the temperature $T_1$ of the relevant parts of the glider, the incoming ambient air temperature $T_0$ and the temperature $T_2$ at the sensor position are functionally related by [Dub11]:

$$\frac{T_2 - T_1}{T_0 - T_1} = \exp\left(-\frac{\alpha \cdot A}{\dot{m} \cdot c_p}\right) \tag{3}$$

The parameters $\alpha$ (heat-transfer coefficient in W/m²K), $A$ (surface of the air-flow channel in m²), $\dot{m}$ (air-mass flow in kg/s) and $c_p$ (specific heat capacity of air in J/kgK) are assumed to be sufficiently constant. Therefore the exponential function can be simplified to the constant $(1-f)$ with $0 \leq f \leq 1$. The value for $f$ must later be determined by a measurement data analysis. Eq. (3) can thus be transformed into

$$T_2 = T_1 \cdot f + T_0 \cdot (1-f) \tag{4}$$

Equating the definition of the heat flow

$$\dot{Q} = k \cdot A \cdot (T_0 - T_1) \tag{5}$$

with the definition of the heat capacity, here derived by time,

$$\dot{Q} = c \cdot m \cdot \dot{T}_1 \tag{6}$$

results in a differential equation of the first order, which describes the temperature profile $T_1$ of aircraft and equipment material as a function of the ambient air temperature $T_0$:

$$\frac{c \cdot m}{k \cdot A} \cdot \dot{T}_1 + T_1 = T_0 \tag{7}$$

$T_1$ follows $T_0$ time-delayed like a first-order low-pass filter with the time constant $\tau = (c \cdot m)/(k \cdot A)$.

As in (3), the parameters $c$ (specific heat capacity of aircraft and measuring equipment in J/kgK), $m$ (mass of the relevant material in kg), $k$ (heat-transfer coefficient in W/m²K) and $A$ (relevant surface of the aircraft in m²) are sufficiently constant so that the time constant $\tau$ also remains constant over the various operating states and altitudes of the aircraft. Also $\tau$ must later be determined by a measurement data analysis.

**Fig. 5** shows the relationship between the environmental temperature $T_0$ in front of the aircraft's nose, the body temperature $T_1$ and the temperature $T_2$ at the sensor position in a block diagram, based on (4) and (7).

**Correction algorithm**

The measuring equipment records $T_2$ with a fixed sample rate of $\Delta t$ = 0.5 sec. Therefore the block diagram in Fig. 5 can be merged in a simple difference equation, with k-1 and k representing the discrete samples of the temperature values:

$$T_{2(k)} = T_{0(k)} + f \cdot (T_{1(k)} - T_{0(k)}), \quad with \quad T_{1(k)} = T_{1(k-1)} + \frac{\Delta t}{\tau + \Delta t} \cdot (T_{0(k)} - T_{1(k-1)}) \qquad (8)$$

In our case a time constant $\tau$ = 80 sec for the low-pass filter and f = 0.75 have been determined in lab experiments and have been verified by correlation of nearly 100 hours of flight data records.

Not shown in Fig 5 is a second time constant of approximately 5 sec reflecting the package and the fixture of the BME280 sensor itself. However, it turns out that the second time constant can be safely ignored as it has only a minor impact on the calculation. This simplifies the correction algorithm.

The recorded data is post-processed such that the temperature $T_0$ in the undisturbed air immediately in front of the glider's nose is available for subsequent calculation by a simple mathematical inversion of (8). It is important to note that the applied data correction for $T_2$ does not depend on the altitude nor does it imply any form of dry-adiabatic lapse rate.

The temperature error can be estimated by comparing a simulated measurement temperature based on a dry adiabatic lapse rate with the actual measurement temperature. The parameters of the simulation model are tuned to achieve maximum correlation between simulation and measurements for all flights and all phases. The accuracy of the calculated temperature has been demonstrated to typically lie within a range of less than +/- 0.1 K of the true value.

If the measured temperature data $T_2$ (the uncorrected raw data of the sensor) is plotted in a diagram with the height above the temperature, a quite characteristic egg-shaped curve results. When climbing, the aircraft can not cool down as fast as the ambient air. It warms the measurement air. The measurement temperature seems to be warmer than the ambient temperature. When descending, it is the other way around. If one compares the temperatures during the descent and the climb at the same altitude, one could mistakenly believe that there is a temperature advantage of the thermals compared to the ambient air as already explained in Fig 3. This underlines the need for the correction method described above.

# Humidity correction

The in-flight measurements show that the humidity sensor of the BME280 exhibits a "dry offset". That means that the relative humidity is always too low resulting in a calculated dew point temperature which is ~ 2.5 °C lower than expected. This dry offset can only be observed in flight. When standing still on the ground or in the lab the dew point values correspond to the values published by the German Weather Service (DWD) for that place and time. We assume the dry offset has to do with the speed at which the air passes by the sensor. The BME280 may not have been designed for such conditions although the data sheet is not explicit about that.

One possible and plausible explanation for the dry offset behavior under those conditions can be related to the measurement principle of the sensor. A thin layer of water builds on the surface of such solid-state capacitive moisture sensors [And94]. This water film is a few molecule diameters thick and caused by adsorption. The thickness of the adsorbed water film depends on the relative humidity and the temperature of the measured air and represents the measurement principle of this sensor type.

As pointed out in [And94] the strong air flow around the sensor may cause a reduction in thickness of the adsorbed water film. This leads to the afore mentioned dry offset and the lower than expected humidity values from the sensor.

Considering Van-Der-Waals-Forces we correct the thickness of the water layer thus correcting the relative humidity values. As with the temperature correction also the humidity correction is independent of the particular flight or any other parameter, so that the same algorithm can be applied to correct all acquired data.

## Results

### First thermal after take-off

**Fig 6** shows the results of a flight on May 14, 2018. The plot uses a format which is widely used for thermodynamics diagrams. The black line shows the uncorrected temperature values exhibiting the afore mentioned loop form ($T_2$ in Fig 4). The red curve on the right shows the corrected temperature ($T_0$ in Fig 4). The blue curve shows the dew-point temperature. Letters denote significant time points during the flight which include: the end of the tow phase after takeoff, searching and circling in the first thermal, cruising to the next thermal. The embedded graphic displays the different phases of the flight, plus an additional curve showing the altitude above takeoff level.

Apparently the corrected temperature curve aligns on a common line for all phases of the flight. There is no measurable temperature difference between the thermal and the ambient air. Furthermore it is obvious that the dew point temperature correlates nicely with the line of constant mixing ratio during the ascent inside the thermal (section C-D-E in Fig 6). The mixing ratio is almost independent of the altitude which lets us conclude that there was no significant mix with (drier) ambient air. Any form of entrainment or dilution would have lead to a change of the mixing ratio with increasing altitude.

During the cruising phase we measure mostly the humidity of the ambient air, clearly drier than inside the thermal (section E-F). But also during that phase of the flight we experience patches of humid air caused by small thermals in between the main thermals which the pilot decided to use. During the tow phase and at the release of the tow-rope (section A-B) we experience temperature spikes. This is probably related to the exhaust from the tow plane's combustion engine.

**Another typical thermal updraft**

**Fig 7** shows another typical updraft, flown on July 25, 2018. Before entering the updraft, inside the updraft and after leaving the updraft, the air temperature in the respective height is always the same. This updraft is not warmer than the ambient air.

Within the thermal, the dew point temperature follows the line of equal mixing ratio. Before entering the thermal and after leaving the thermal, the air is clearly drier. It can be assumed that the dew-point temperature of the ambient air is approximately on the dotted blue line. That is, the air in the updraft has a lower density than the ambient air because it contains more humidity, not because it is warmer. The driving force behind the buoyancy is the moisture difference, not the temperature difference!

A closer look at the lapse rate of the thermal shows that it is slightly warmer than the dry adiabat. The red temperature curve is not parallel to the gray dry adiabats. This is another more indirect proof that the temperature inside the updraft corresponds to the environment.

**Lateral entrainment**

The dew-point temperature in **Fig 8** follows almost constantly a mixing ratio of approximately 8.5 g/kg. In fact the humidity inside the thermal becomes a little drier with increasing height. The mixing ratio moves a little to the left. This change indicates lateral entrainment at the edges of the updraft with an entrainment rate of approximately 0.3 per km. So lateral entrainment seems to be relatively small in thermal updrafts.

**Circling up to cloud base**

The measurement data are so good that they also allow a look at some interesting details. **Fig 9** shows a flight up to the cloud base, where the glider flew in an area of incipient condensation. Good to see is the cooling of the ambient air under the cloud after the aircraft has left the updraft stream. This is because water droplets fall out of the cloud, evaporate and thereby cool the air.

In close vicinity to the updraft air flow, downdrafts are noticeable. They are relatively dry. These downdrafts are created by entrainment on cloud top, sink through the entire cloud and exit out of the cloud base [Pre18]. Of course, these downdrafts, like the updrafts, have the same temperature as the ambient air. They are neither warmer nor colder. They only fall because they are drier and thus have a higher density than the ambient air.

**Blue thermals with different strength and estimated mixing at ground level**

Two consecutive blue thermals A and B in a distance of approximately 6 km are analyzed in **Fig 10**. Both updrafts have the same temperature. But it can be seen that the stronger thermal with 2.3 m/s is moister compared to the ambient air than his weak neighbor with only 1.2 m/s.

We have only few data points for thermals in low altitude (< 500m GND). Hence we can't be really sure what the temperature curves look like between the surface layer and the first few hundred meters height. However, we believe there is a positive temperature difference which is needed to get the thermal off the ground. Down there and only down there thermals are really warmer than the environment. As the air climbs this temperature difference decreases by the ascend itself as well as by mixing with cooler ambient air. As the formation of an updraft is connected with horizontal winds near the ground, intense turbulent mixing will soon reduce any temperature advance. Thus there seems to be higher entrainment close to the ground compared to higher areas.

## Conclusions

Difference in air-density is the driving force behind thermals. Difference in temperature and humidity is the primary cause of this buoyancy. Extensive measurements over Germany in the summer of 2018 show that humidity is the dominating moving force of the thermals, at least in the upper three quarters between ground and cloud base. In order to recognize this dominance, it is essential either to prevent interfering influences on the temperature measurement or to compensate the measured data mathematically. A faulty temperature measurement can easily lead to the conclusion that the drive of the thermals is wrongly attributed to a, actually not existing, temperature difference. Furthermore, the algorithms for calculating the thermal strength should also take into account the moisture difference and not solely rely on temperature difference. Future measurements will investigate more closely the lower realms where temperature difference is expected to dominate the convective updraft. As the temperature difference disappears within a few hundred meters due to lateral entrainment and mixing a thermal's ascent is non-adiabatic. Maybe another word for convective updrafts should be used: They are obviously more a "moistal" than a "thermal".

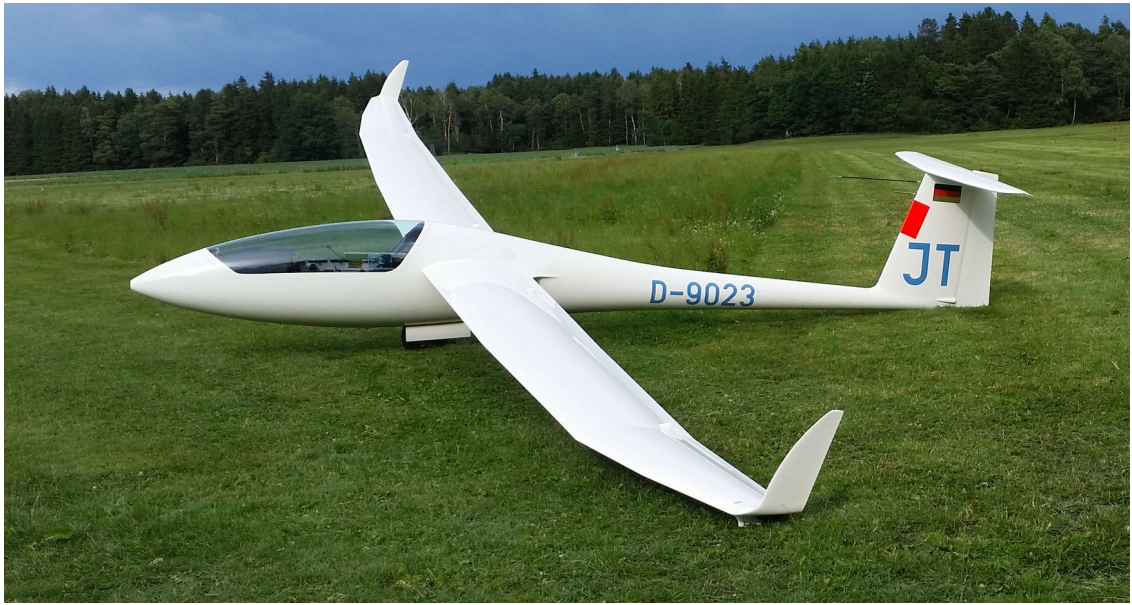

**Fig 1**: The Ventus-2 glider used for environmental measurements.

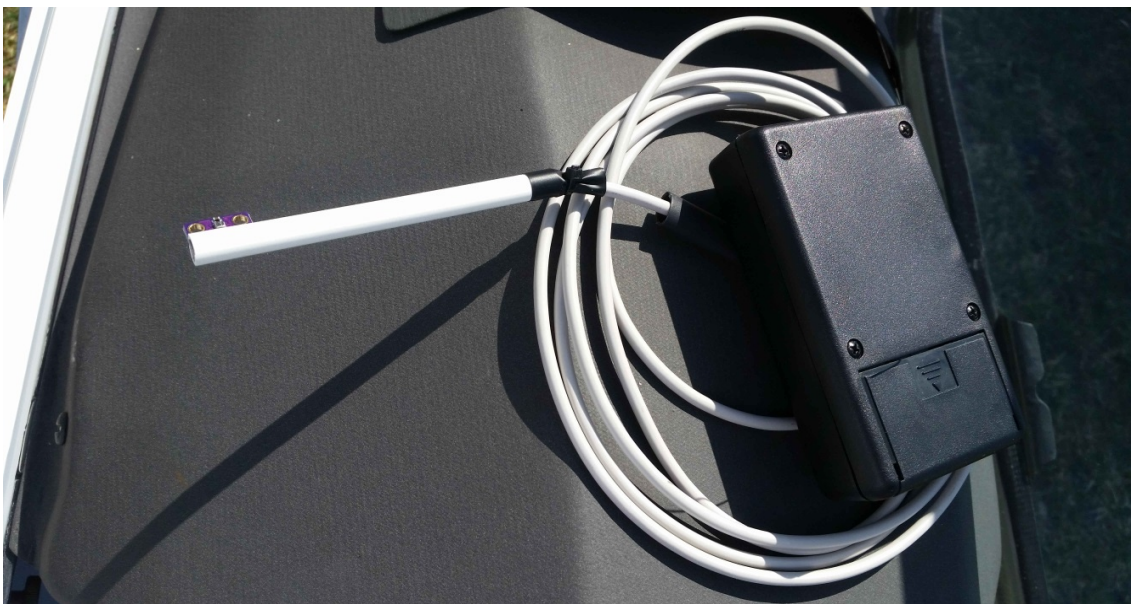

**Fig 2**: Measurement hardware installed on the Ventus-2 glider.

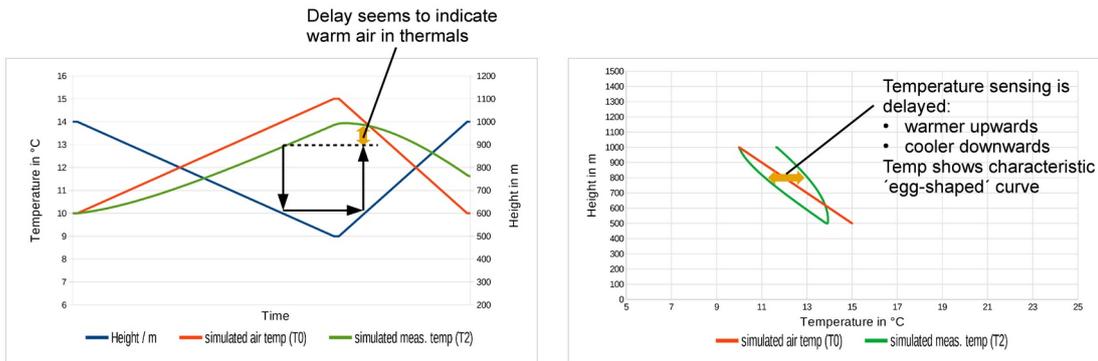

**Fig 3:** The uncorrected raw temperature sensor signal indicates a faulty temperature advance of a thermal.

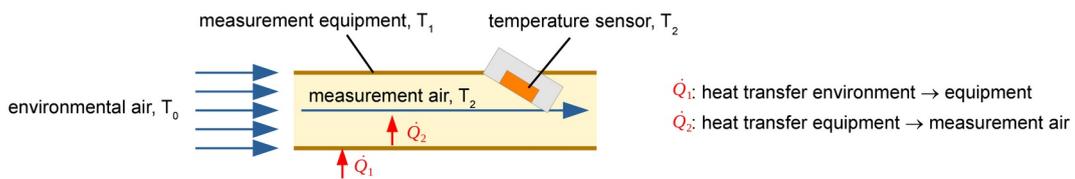

**Fig 4:** The influence of heat transfer on the measured air temperature.

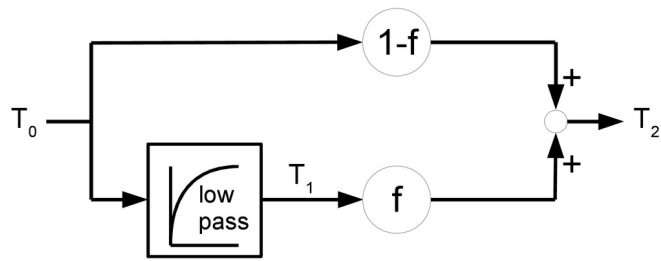

**Fig 5:** Block diagram to describe the influence of heat transfer and heat capacity on the measured air temperature with $T_0$: environmental air temperature in front of the aircraft; $T_1$: temperature of the equipment; $T_2$: air temperature at the sensor position.

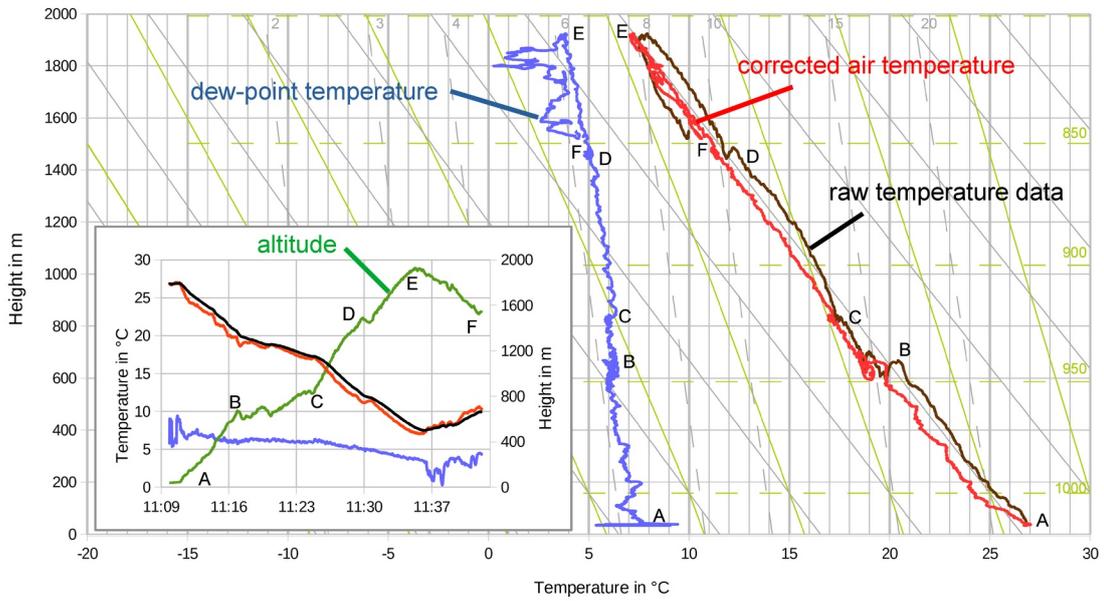

**Fig 6:** Temperature and dew-point temperature during a flight on May 14, 2018.
A: takeoff; B: release; C: entering 1st thermal; D: entering 2nd thermal;
E: leaving thermal; F: entering 3rd thermal

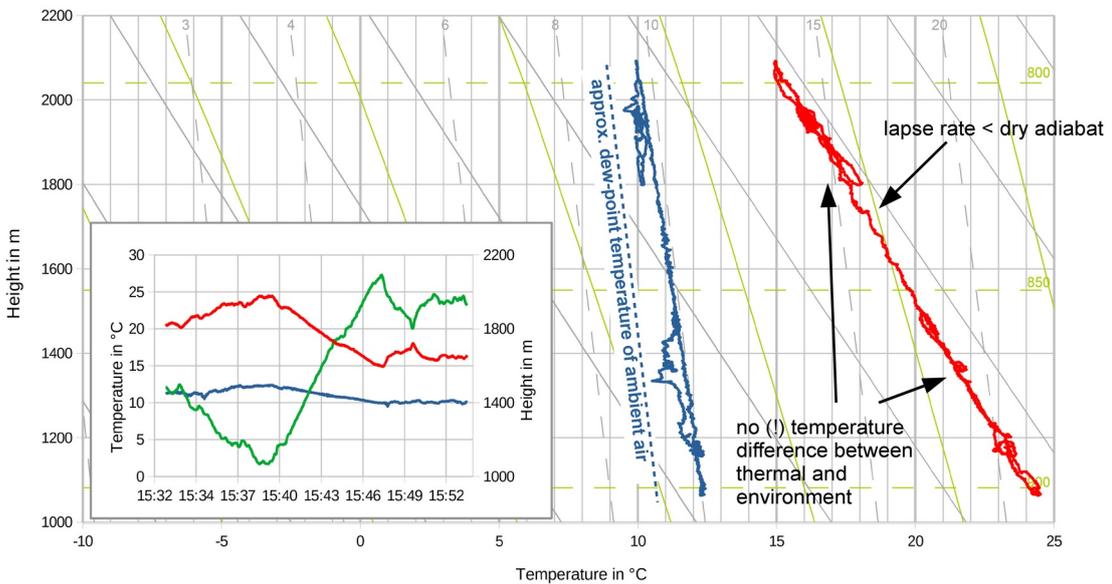

**Fig 7:** Temperature and dew-point temperature during a flight on July 25, 2018.

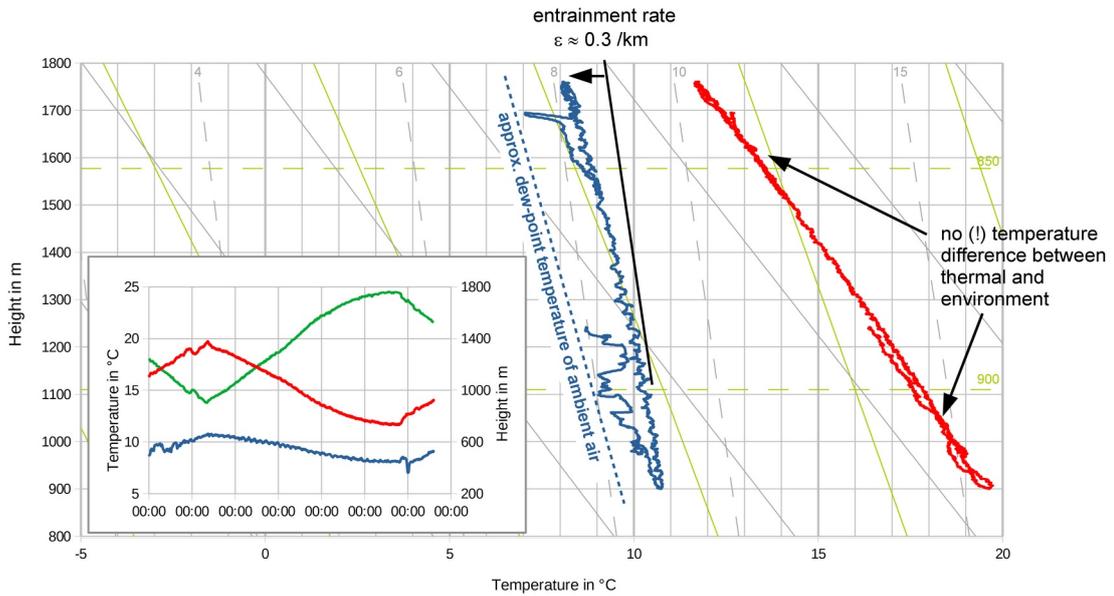

**Fig 8:** Temperature and dew-point temperature during a flight on May 26, 2018 with typical lateral entrainment.

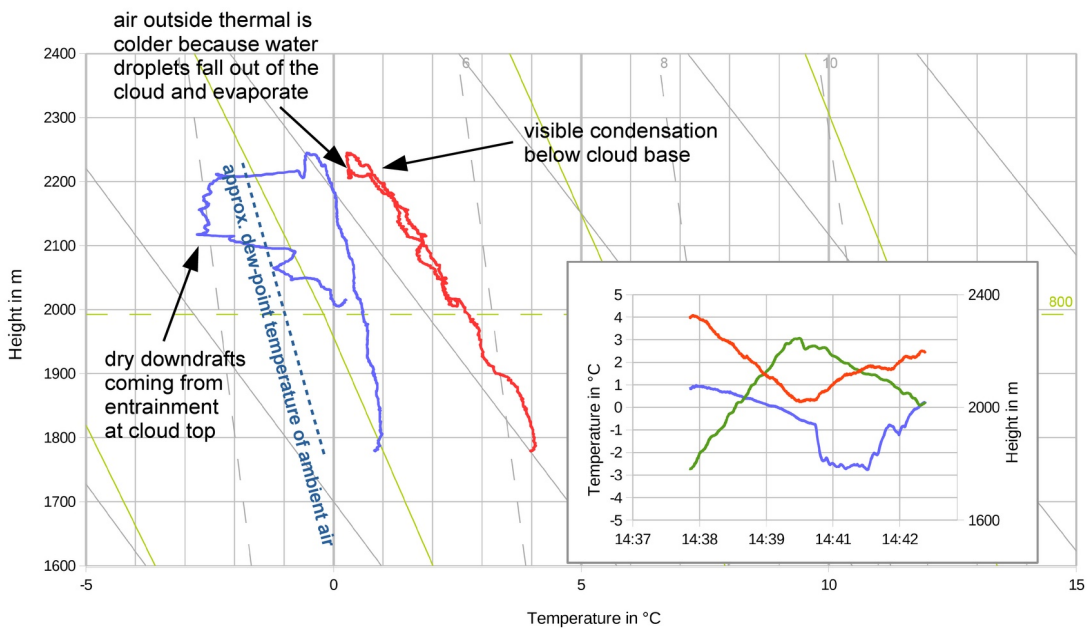

**Fig 9:** Temperature and dew-point temperature at cloud base during a flight on August 26, 2018.

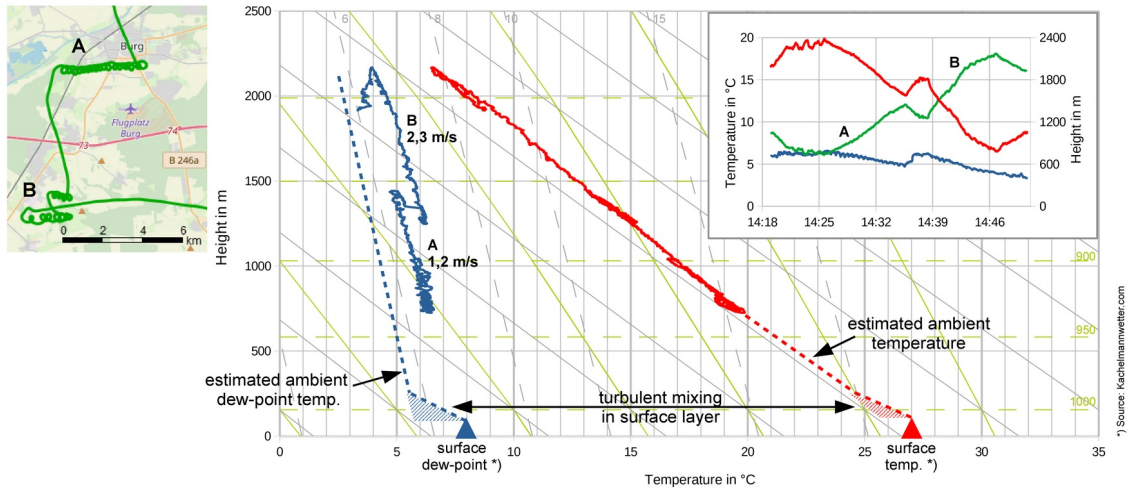

**Fig 10:** Two consecutive blue thermals A and B with different updraft velocities during a flight on May 13, 2018.